\documentclass[aps,prb,preprint,superscriptsaddress,showpacs,amssymb]{revtex4}
\usepackage{graphicx}
\usepackage{dcolumn}
\usepackage{bm}

\begin{document}


\title{Electronic Restructuring in Shape-Memory Alloys: Thermodynamic and electronic structure studies of the
martensitic transition}

\author{J. C. Lashley}
\affiliation{Los Alamos National Laboratory, Los Alamos, NM 87545,
USA}
\author{P. S. Riseborough}
\affiliation{Physics Department,
   Temple University,
   Philadelphia, PA 19122, USA}
\author{C. P. Opeil}
\affiliation{Department of Physics, Boston College, Chestnut Hill,
MA 02467, USA}
\author{R.~K. Schulze}
\affiliation{Los Alamos National Laboratory, Los Alamos, NM 87545,
USA}
\author{B. Mihaila}
\affiliation{Los Alamos National Laboratory, Los Alamos, NM 87545,
USA}
\author{W. L. Hults}
\affiliation{Los Alamos National Laboratory, Los Alamos, NM 87545,
USA}
\author{J. C. Cooley}
\affiliation{Los Alamos National Laboratory, Los Alamos, NM 87545,
USA}
\author{J. L. Smith}
\affiliation{Los Alamos National Laboratory, Los Alamos, NM 87545,
USA}
\author{R. A. Fisher}
\affiliation{Lawrence Berkeley National Laboratory, Berkeley,
California 94720, USA}
\author {O. Svitelskiy}
\affiliation{National High Magnetic Field Laboratory, Florida State
University, Tallahassee, FL 32310, USA}
\author {A. Suslov}
\affiliation{National High Magnetic Field Laboratory, Florida State
University, Tallahassee, FL 32310, USA}

\author{T. R. Finlayson}
\affiliation{School of Physics, Monash University, Clayton,
Victoria, Australia 3800}

\date{\today}

\begin{abstract}
Using a variety of thermodynamic measurements made in magnetic
fields, we show evidence that the diffusionless transition (DT) in
many shape-memory alloys is related to significant changes in the
electronic structure. We investigate three alloys that show the
shape-memory effect (In-24 at.\% Tl, AuZn, and U-26 at.\% Nb). We
observe that the DT is significantly altered in these alloys by the
application of a magnetic field. Specifically, the DT in InTl-24
at.\% shows a decrease in the DT temperature with increasing
magnetic field. Further investigations of AuZn were performed using
an ultrasonic pulse-echo technique in magnetic fields up to 45~T.
Quantum oscillations in the speed of the longitudinal sound waves
propagating in the [110] direction indicated a strong acoustic de
Haas-van Alphen-type effect and give information about part of the
Fermi surface.
\end{abstract}


\maketitle

\section{Introduction}

Martensitic transitions (MTs) are diffusionless structural
transitions that generally occur between a high-temperature cubic
phase and a lower-temperature phase with a lower symmetry. The MT
proceeds via an atomic rearrangement that involves a collective
shear displacement and shows almost no change in volume. Often the
MT can be preceded by a higher temperature pre-MT. According to
Friedel\cite{Friedel 1974}, the martensitic transition is entropy
driven. In Friedel's picture, the high-temperature cubic structure
is stabilized by the entropy of low-frequency phonon modes, and the
low-temperature close-packed structure is energetically stable due
to the larger coordination number. Measurements of the elastic
constants\cite{Guenin 1977,Planes01} and the phonon-dispersion
relations\cite{Mori 1975,Nagasawa93,Planes01} show the existence of
anomalies in the high-temperature phases. In these bcc phases, the
transverse acoustic phonon frequencies are observed to
soften\cite{Petry 1991,Manosa93a} as the temperature is reduced
towards the MT temperature, but generally the softening is
incomplete since it is arrested by the MT before the phonon
frequency reaches zero\cite{Shapiro 1989,Planes96}. Presumably the
incomplete softening is a result of anharmonic interactions, which
would also be in accord with the transition being of first-order and
not continuous. The observed partial softening can be used to infer
the symmetry of the low-temperature structure\cite{Ho 1984,Stassis
1978,Guenin80}. The entropy deduced from the phonon density of
states for most shape-memory alloys has been found to account for
65\% to 75\% of the total entropy change at the transition, and can
be as high as 90\% in Cu-based shape-memory alloys\cite{Manosa93b}.

The above picture has led to theoretical models\cite{Gooding
1988,Brown 2002} that describe the transition entirely in terms of
entropy and anharmonic lattice dynamics. However the entropy-driven
scenario is deficient because the explanation does not explicitly
address the energetic stability of the close-packed phases at
low-temperature. It has been found that it is the difference in the
electronic energies that dominate the structural stability energy of
transition metals\cite{Pettifor}. The above statement is also true
for simple metals\cite{Cohen,Moriarty} since the lattice or Madelung
energies of the body centered and the close-packed structures are
extremely close\cite{Grifalco,Hafner}. In fact, it has long been
known that in Hume-Rothery alloys, the electron concentration per
atom is extremely important in determining which structure is
stable\cite{HumeRothery 1954}. Shortly after Hume-Rothery's
discovery, Jones suggested that the stability\cite{Jones 1937} of
the crystal structure could be understood in terms of an energy
lowering due to the Fermi surface being close to the active
Brillouin zone boundaries. Although Jones's simple one-electron
picture has been shown to be deficient\cite{Heine and Weaire 1970},
it still provides a remarkably good phenomenological description of
these materials. Because many MTs proceed by shuffles (that is
displacements involving quasi-static phonons with (fractional)
commensurate wave vectors with uniform shears\cite{Burgers}), it is
possible that the electron concentration could also play an
important role in producing the phonon anomalies of martensitic
Hume-Rothery alloys. Recently, there have been an increasing number
of investigations that emphasize the importance of the electronic
structure. Transport measurements have shown that significant
changes occur in the magnetoresistance of AuZn alloys, which imply
that significant changes in the Fermi surface have taken
place\cite{McDonald 2005}. De Haas-van Alphen
measurements\cite{Goddard 2005} made above and below the MT
temperature showed drastic reconstruction of the Fermi surface,
thereby confirming the conclusions of McDonald \emph{et
al.}\cite{McDonald 2005}. Electronic-structure
calculations\cite{Zhao 1992} indicate that Fermi-surface nesting may
also be responsible for the phonon softening observed\cite{Satija
1984} in the pre-martensitic phases of NiTi, NiAl, and AuCd alloys,
as well as that predicted for vanadium and niobium at high
pressure\cite{Landa}. The electronic momentum distribution has been
measured for a NiAl alloy by Compton-scattering
experiments\cite{Dugdale 2006}, which allowed the Fermi surface to
be mapped out and which shows indications of nesting in agreement
with the theoretical predictions of Zhao and Harmon\cite{Zhao 1992}.

In this paper, we report thermodynamic measurements in magnetic
fields on three nonmagnetic alloys that show the shape-memory
effect. These measurements demonstrate that magnetic-field-induced
changes of the electronic structure strongly couple to the DT.
Acoustic de Haas-van Alphen measurements, which are a direct probe
of the Fermi surface, on AuZn using pulse-echo sound-speed
measurements in high-magnetic fields provide further evidence of the
important role of the conduction electrons in the DT. We show in
AuZn that the velocities of the longitudinal acoustic phonons depend
on the magnetic field and that the phonon frequencies exhibit an
oscillatory dependence on the inverse of the magnetic field. This
phenomenon is understood\cite{Shoenberg's book} as a consequence of
successive Landau tubes sweeping through the Fermi surface, which
modulate the dielectric constant at the reciprocal lattice vectors
and, through the screening of the pair potential, modulate the
observed phonon velocities. This interpretation is in accord with
the observation\cite{finlayson_smith} that (in InTl alloys) Kohn
anomalies in the phonon-dispersion relations occur at wave vectors
almost commensurate with reciprocal lattice vectors.

\section{Thermal-expansivity measurements}

The coefficient of linear thermal expansion, $\alpha$, was measured
in a three-terminal capacitive dilatometer~\cite{George} over the
range $100~K\leq T\leq 300~K$. The specimen was held against a rigid
fixed platform on the bottom, and the top was in contact with a
spring-loaded lower capacitor plate, which was located below an
upper capacitor plate. As the specimen expanded and contracted, it
changed the size of the gap $D$ between the capacitor plates. It
should be noted that there is an unavoidable constant stress to the
sample from the spring connecting to the lower capacitor plate. This
stress is the same from measurement to measurement. Because the
temperature dependence of this capacitance includes contributions
from the specimen and the cell, the cell effect is measured
separately and subtracted. Capacitance was measured with a
Andeen-Hagerling 2500~A 1~kHz ultra-precision capacitance bridge.
Data were recorded on cooling at a rate of 0.2~K/min, and the bridge
was set to take 10 averages at each point.

\begin{figure}[t]
    \centering \includegraphics[width=\columnwidth]{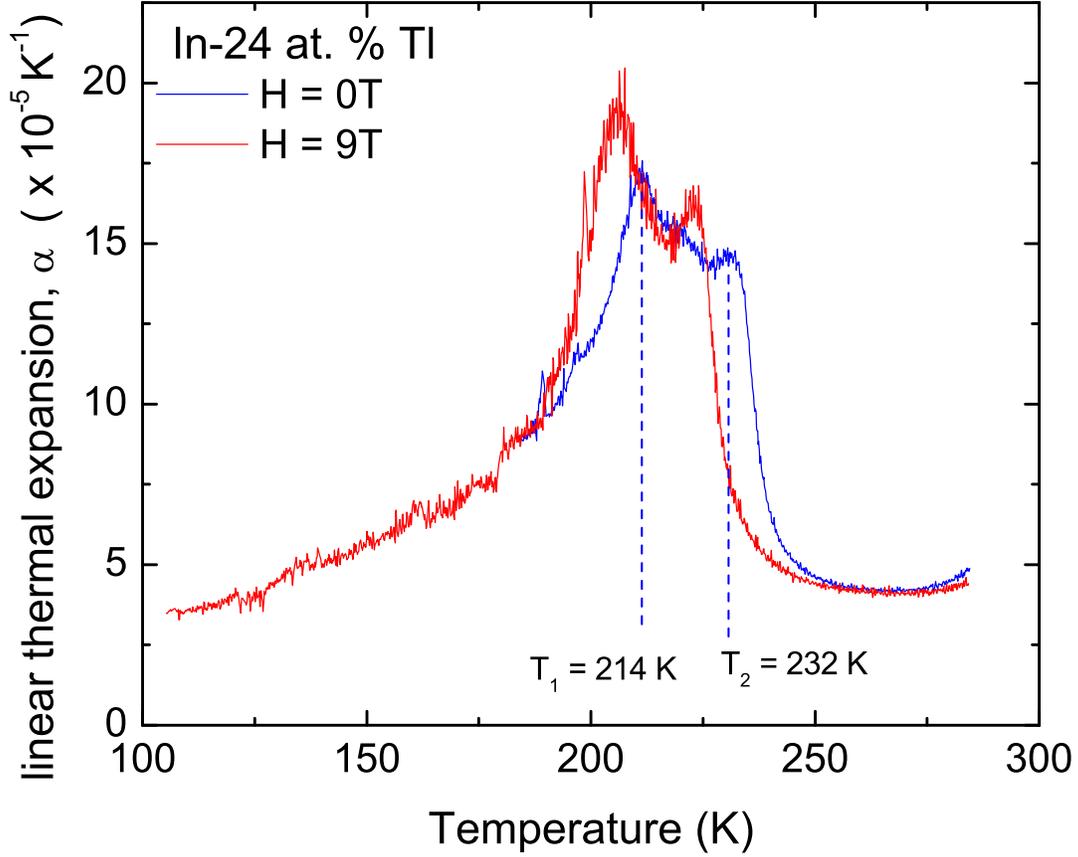}
    \caption{Magnetic field dependence of the
    coefficient of linear thermal expansion for In-24 at.\% Tl in the vicinity of the
    MT. The magnetic field is applied at room temperature parallel to the [010] crystallographic axis of the fcc phase.
    In both cases the data were recorded on cooling from 300~K at a rate of
    0.2~K/min.}    \label{Fig1}
\end{figure}

\begin{figure}[t]
    \centering \includegraphics[width=\columnwidth]{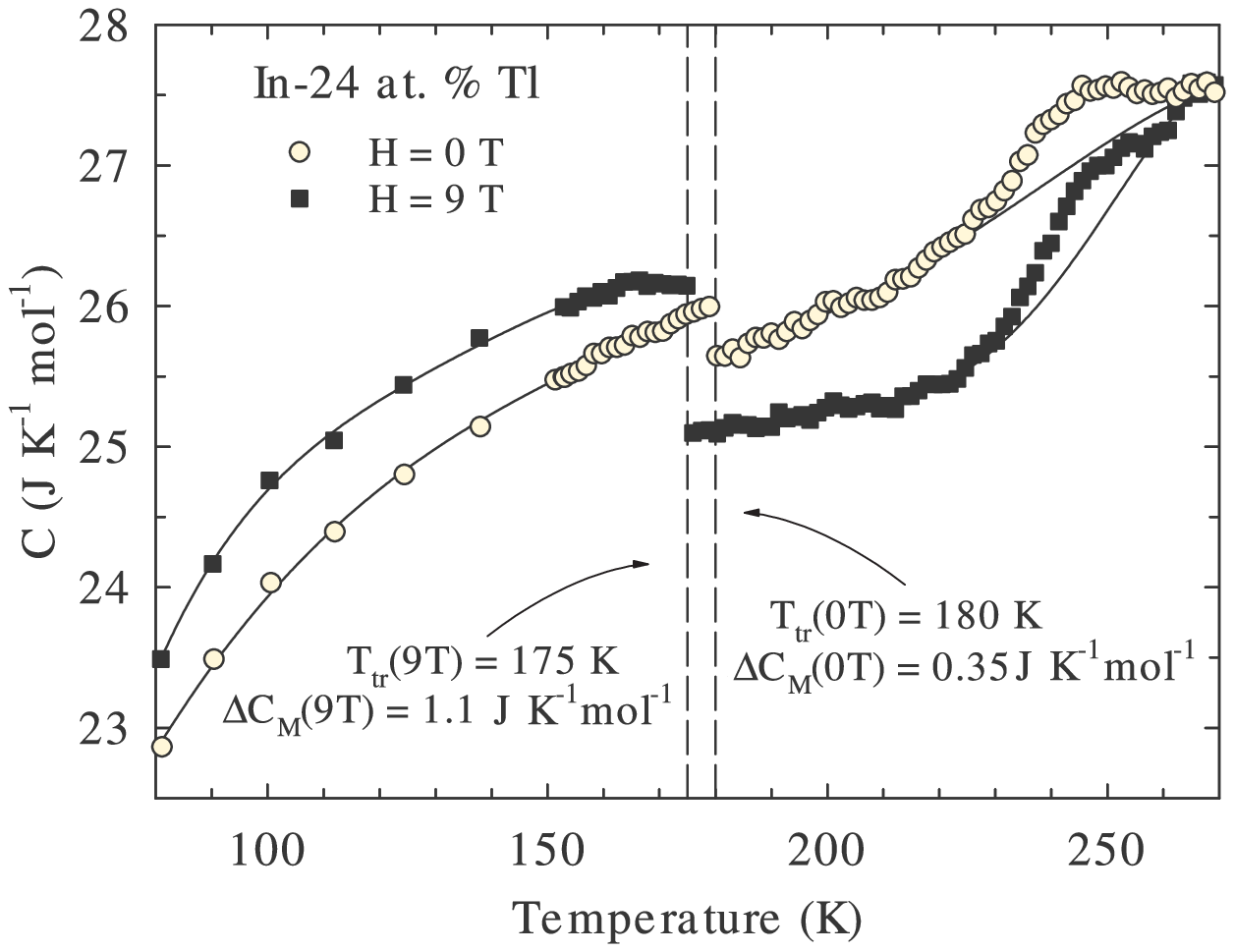}
    \caption{Magnetic field dependence of the
    specific heat of In-24 at.\% Tl in the vicinity of MT.
    The magnetic field is applied parallel to the [100] crystallographic axis of the fcc phase.
    Values for the transition temperatures, $T_{tr}$, and the
    discontinuities in the specific heat evaluated at the MT, $\Delta$$C_M$
    are shown in the plot. In the absence of the magnetic field the transition
    occurs at $T_{tr}(0T)$=180~K, whereas for H=9~T the transition temperature
    becomes $T_{tr}(9T)$=175~K.
    }    \label{Fig2}
\end{figure}

The performance of the cell was tested against an aluminum standard,
and the measurements fell within 1\% of the reference material. With
this apparatus, the operative equation to obtain the thermal
expansion is
\begin{equation}
\ \alpha  = \frac{1} {L}\frac{d} {{dT}}\left[ {D_c  - D_s } \right]
+ \alpha _{Cu}  . \
\end{equation}
Here, $L$, denotes specimen length at room temperature, $D_c$, the
gap from the cell effect, $D_s$, the gap from the expansion or
contraction of the specimen, and, $\alpha_{Cu}$, the coefficient of
linear thermal expansion for copper. The sample was a single crystal
of an InTl alloy with 24 at. \% Tl which is a composition in the
range where the high-temperature fcc phase becomes unstable to an
fct phase at lower temperatures\cite{Meyerhoff}. Following the
standard explanation of the tetragonal distortion\cite{Heine2}, one
expects that the fct phase is stabilized by large gaps in the
electronic dispersion relations at the Bragg planes derived from the
\{111\} and \{200\} reciprocal lattice vectors of the fcc structure.
Since the {\it c/a} ratio is greater than unity in the tetragonally
distorted phase\cite{Guttman}, it may also be expected that the
(200) and (020) reciprocal lattice vectors should move closer to
$2k_F$ in the tetragonal phase, while (002) moves in the opposite
direction. The results of the thermal-expansivity measurements on an
InTl alloy with 24 at. \% Tl are shown in Fig.~\ref{Fig1}. The
sample was oriented with a [100] axis of the fcc structure parallel
to the axis of the capacitor. Thermal- expansivity measurements were
performed with decreasing temperature, once with zero field and once
with a field of H = 9~T oriented parallel to the [010] direction of
the fcc structure. Two peaks are evident in both the H = 0~T and 9~T
data, and the peak structure spans 180~K to 260~K in the H = 0~T
data. If one assigns characteristic temperatures to each of these
two peaks ($T_1$ = 214~K and $T_2$ = 232~K), one sees a that the
temperature is reduced ($T_1$ = 205~K and $T_2$ = 224~K) in 9~T.
Both measurements were repeated on the same crystal, and the peak
temperatures did not shift; hence there is no evidence of cycling.
With H = 9~T, $T_1$ is reduced by approximately 6~K and $T_2$ is
reduced by 8~K. The magnitude of the decrease in the characteristic
temperatures roughly correspond to the magnetic-energy scale of an
electron in the applied field ($k_B \Delta T_1 \approx \mu_B H$),
which implies that the transition intimately involves the conduction
electrons. Previous zero-field results of Liu \emph{ et
al.}\cite{Liu93} show that the slight compression along the [100]
direction causes the crystal to transform so that [100] axis remains
an {\it a-}axis of the tetragonal structure. The expansivity data
shows a less pronounced field dependence when the field is applied
in the fcc [001] direction, which is consistent with tetragonal
anisotropy. We tentatively identify the [001] direction with the
{\it c-}axis by assuming that the lowering of the structural energy
is primarily due to electrons near the Bragg planes closer to the
Fermi surface\cite{Ashcroft} since the Lorentz force from fields
perpendicular to the planes should be more effective in changing the
electronic structure. Larger changes in electronic binding energy
are expected to result in larger changes in the transition
temperature.

The reason for the two-peak structure in In-24 at.\% Tl is a result
of the twinning structure that accompanies the martensitic
transition. Video evidence of the development of the transition
morphology through the transformation temperature regime for an
In-19 at.\%Tl alloy shows that over an approximate 25~K temperature
range in the vicinity of the martensitic transition, the surface
bands appear discontinuously and also thicken as the temperature is
decreased\cite{video}. In addition, at a temperature below the MT
one set of variants will disappear at the expense of another set,
giving rise to the double peak structure. InTl is a soft alloy
($\Theta_D$ = 94~K), and the spring-loaded capacitor imparts a small
amount of compressive stress on the sample. This stress then
controls the allowed variant morphology accompanying the transition.
It should be noted that the anomaly in the measured linear thermal
expansion coefficient with temperature is consistent with the
measurement direction being predominantly an {\it a-}axis following
the cubic-to-tetragonal transition. Comparable behavior has been
observed for the anomaly in $\alpha (T)$ measured by capacitance
dilatometry by Liu and coworkers\cite{Liu93}, who also showed that
by applying a biaxial stress field to a [100] crystal, perpendicular
to the measurement direction, the observed result for the anomaly in
$\alpha (T)$ became {\it c-}axis-like.

\section{Specific-heat measurements}

Specific-heat measurements were made on single crystal samples of
In-24 at.\% Tl, taken from the same batch that was used for the
linear-expansion measurements, and polycrystalline samples of the
U-26 at.\% Nb alloy. Figure~\ref{Fig2} shows the results for the
specific heat in the vicinity of the MT, measured using a
thermal-relaxation method in magnetic fields applied along the [100]
direction with field strengths between 0 and 9 T. The specimen was
thermally attached to a specimen platform with a thin layer of
Apiezon N grease. The specific heat of the empty sample platform and
the specific heat of Apiezon N were measured separately and
subtracted from the total specific heat to obtain the specific heat
of the sample. The performance of the calorimeter was rigorously
tested with a variety of conditions and materials confirming its
accuracy\cite{Lashley}. In zero magnetic field, anomalous
temperature dependence for the specific heat in decreasing
temperature is first detected at about 260~K. Then at 180~K there is
a clear discontinuity with the specific heat increasing by 0.31
J~K$^{-1}$mol$^{-1}$. The difference in the martensitic transition
temperature between the specific heat and the thermal expansion
could be caused by slight changes in the Tl concentration.
Application of the magnetic field depresses the temperature of this
discontinuity by 5~K, and $\Delta$$C_M$ increases by a factor of
three to 1.1~J~K$^{-1}$mol$^{-1}$ for H = 9~T. The change of the
temperature of the discontinuity with H = 9~T coincides with the
observed changes in the characteristic temperatures found in the
thermal expansion coefficient. When the 9~T field was applied along
another crystallographic axis of the fcc phase, say [001], the
measured discontinuity of the specific heat was reduced which, for
reasons previously discussed, is consistent with the [001] axis
becoming the {\it c-}axis of the tetragonal phase.

\begin{figure}[t]
    \centering \includegraphics[width=\columnwidth]{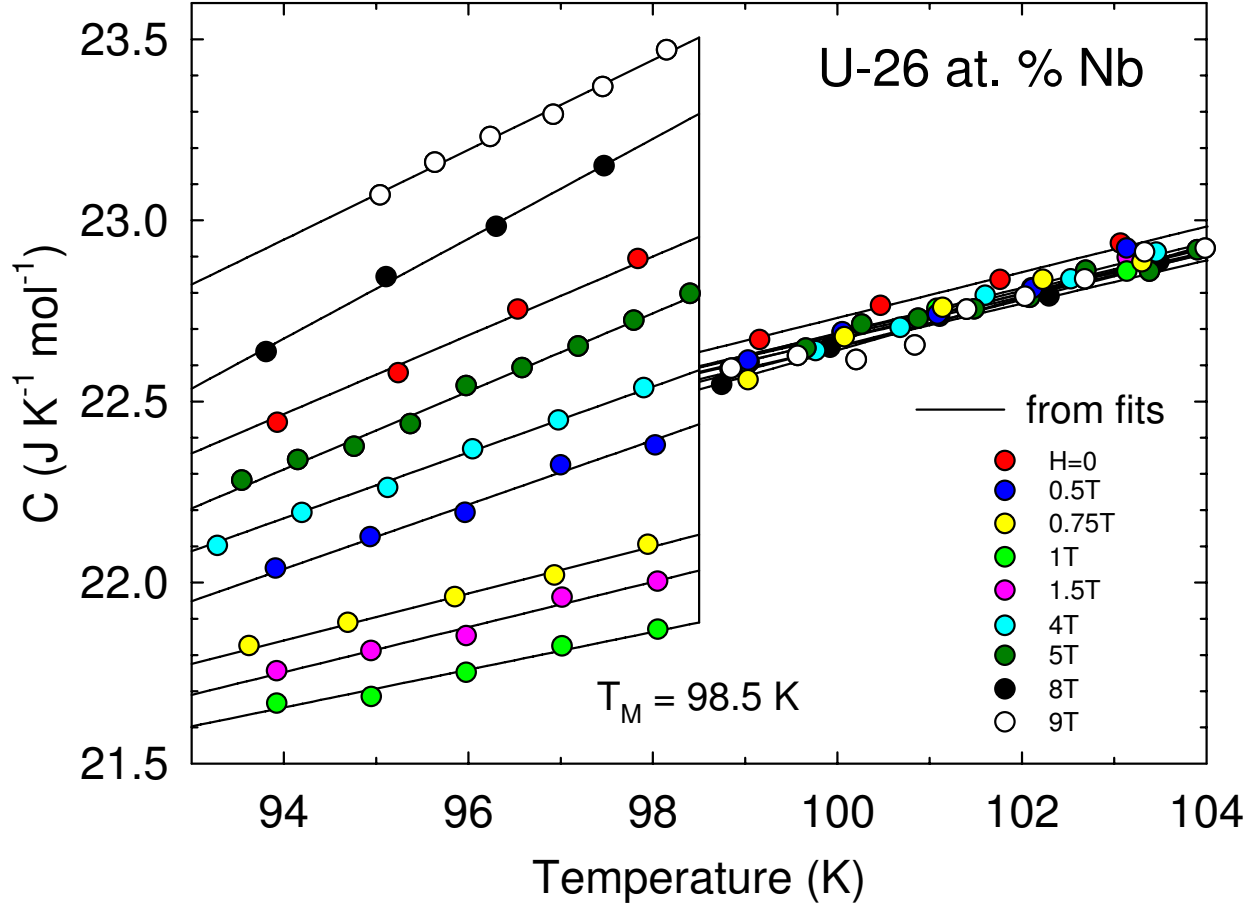}
    \caption{Specific heat data in the vicinity of the MT in U-26 at.\% Nb
    as a function of magnetic field. All data were recorded on warming.}    \label{Fig3}
\end{figure}

Figure~\ref{Fig3} shows the specific-heat data in the vicinity of
the MT in a U-26 at.\% Nb alloy as a function of magnetic field.
 Above the transition temperature, all data are congruent to
within $\pm$~1\%. Linear extrapolations were made over temperatures
spanning the MT and were used to derive $\Delta$$C_M (H)$, which has
an estimated precision of $\sim$~1\%. The negative suppression in
the data for fields in the range $0.3~T \leq H \leq 4~T$ suggests
that the magnetic field is relaxing the lattice strain. Because the
UNb alloys are polycrystalline and have a low-symmetry orthorhombic
structure, microstrain originating from misorientations of grain
boundaries are known to affect the low-temperature specific heat and
high-temperature specific heat of $\alpha$-uranium (also
orthorhombic)~\cite{Mihaila06,Manley}. Above 1~T the field
dependence is indicative of a significant change in the heat
capacity. We have measured such an effect in the shape-memory alloy
AuZn where the opening of a gap at the Fermi surface\cite{McDonald
2005, Goddard 2005} affects the shape of the excess specific heat at
the transition temperature.

\section{Acoustic measurements}

A pulse-echo technique was used to explore the properties of
longitudinal sound waves with wave vectors along the $[110]$
direction of the stoichiometric (1:1) AuZn crystal~\cite{Darling
2002,LT}. These measurements were made in the temperature range
$0.07~K\leq T \leq 50~K$ with magnetic fields that were varied up to
45~T. The samples were mounted on a single-axis goniometer that
allowed the orientation of the sample in the magnetic field to be
controlled to within one degree. The relative change in the speed of
sound was measured with a precision of $10^{-7}$.

\begin{figure}[t]
    \centering \includegraphics[width=\columnwidth]{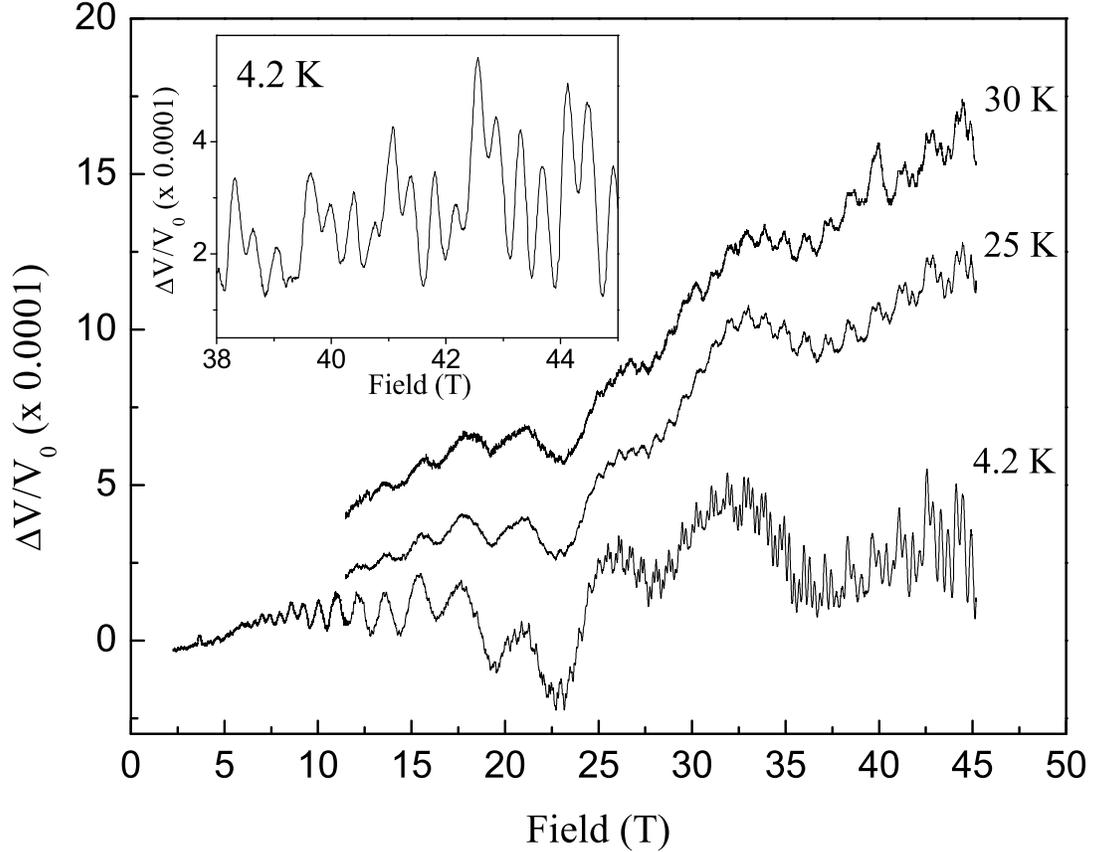}
    \caption{Oscillations of the speed of sound in AuZn measured at several
    temperatures in the martensite phase.
    The inset shows the high-field part of these oscillations.
    The 60.3 MHz longitudinal sound wave and the field are both directed
    along [110] crystallographic direction.}    \label{Fig4}
\end{figure}

\begin{figure}[t]
    \centering \includegraphics[width=\columnwidth]{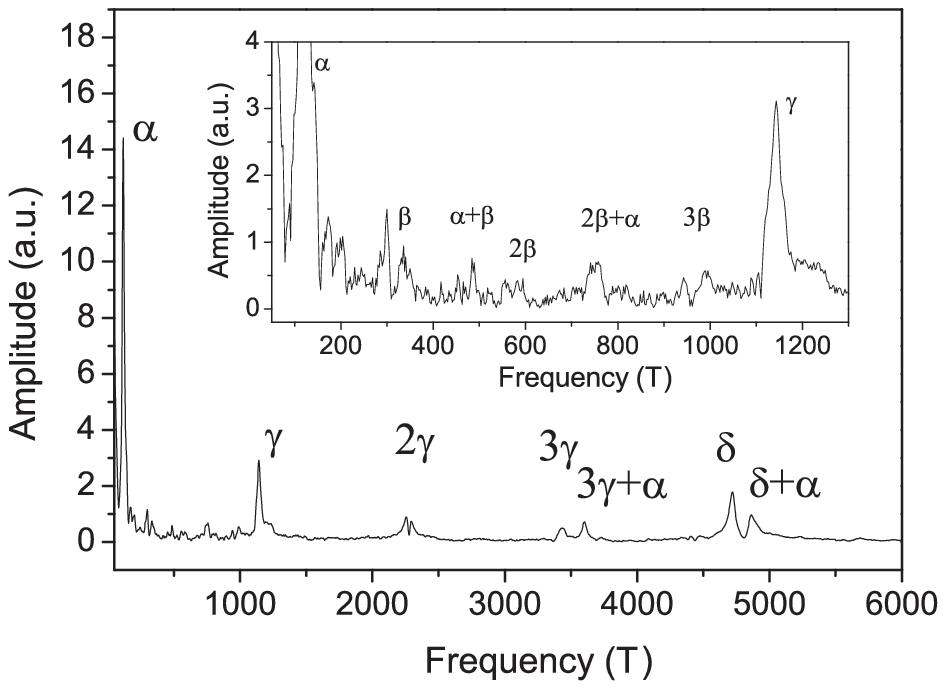}
    \caption{Fourier image of the 4.2~K oscillations (Fig.4).
    The inset shows a magnified view of the low-frequency part of the spectrum.
    Some peaks are split by 46~T, although this frequency was not explicitly detected.}    \label{Fig5}
\end{figure}

In close agreement with the results of Schiltz \emph{et
al}.\cite{Schiltz 1971}, the speed of longitudinal sound in the
$[110]$ direction shows a linear increase with decreasing
temperature between room temperature and the transition temperature
64~K.  The speed of sound exhibits quantum oscillations, shown in
Fig.~\ref{Fig4}. The magneto-acoustic oscillations in AuZn can be
compared with measurements\cite{McDonald 2005, Goddard 2005} of the
de Haas-van Alphen oscillations. Previously, it was found that the
de Haas-van Alphen effect persisted up to 100~K in pulsed magnetic
fields. Since the present study is limited to fields below H = 45 T,
the oscillations are only observable up to a maximum temperature of
$\sim 50~K$. The amplitude of the magneto-acoustic oscillations
reaches the largest values when the magnetic field is oriented along
the $[110]$ axis. The Fourier spectrum shown in Fig.~\ref{Fig5}
exhibits peaks at discrete frequencies. This spectrum shows peaks
$\beta$ at 303~T, $\gamma$ at 1141~T, and $\delta$ at 4770~T, in
accordance with the measurements of the de Haas-van Alphen
effect\cite{McDonald 2005}. In addition, we also observe the
$\alpha$ peak at 120~T, which was not observed in the de Haas-van
Alphen measurements. The oscillations are interpreted in terms of
Landau tubes sweeping through the Fermi surface, which produce
oscillations in the dielectric constant. The dielectric constant is
involved in screening the indirect interactions between the ions and
is responsible for their asymptotic Friedel oscillations with wave
vector $2k_F$ \cite{Morell Cohen,Harrison}. The oscillatory
dependence of the pair-potential on the magnetic field then produces
oscillations of the phonon frequencies. The temperature-dependence
of the amplitudes of the magneto-acoustic oscillations were fit to
the Lifshitz-Kosevitch formula
\begin{equation}
A={{2 \pi^2 m^{*}c k_B T \over e \hbar B}\over\sinh \Bigl ({2 \pi^2
m^{*}c k_B T\over e \hbar B} \Bigr )}  \ \exp \bigg[-{2 \pi^2 m^{*}c
k_B T_D\over e \hbar B}\bigg] \ ,
\end{equation}
where $m^{*}$ is the effective mass, and $T_D$ is the Dingle
temperature. The effective masses of $m_{\gamma} \approx 0.21\pm
0.01$ ($m_e$) and  $m_{\delta} \approx 0.32 \pm 0.01$ ($m_e$) for
the $\gamma$ and $\delta$ orbits respectively are in agreement with
earlier experimental and theoretical predictions for the hexagonal
martensite phase\cite{McDonald 2005}. The effective mass for the
$\alpha$ sheet is determined to be $m_{\alpha} \approx 0.12 \pm
0.01$ ($m_e$). The Dingle temperatures are estimated as $T_{\alpha}
\ \approx 19 \pm 1$~K, while $T_{\gamma} \ \approx 16 \pm 3$~K, and
$T_{\delta} \ \approx 16 \pm 1$~K corresponding to cyclotron
mean-free paths of the order of 600~\AA. This mean-free path is
comparable to that inferred from the measured value of the residual
resistivity which is 6.03~$\mu$$\Omega$cm. The observation of the
$\alpha$ orbit combined with previous de Haas-van Alphen
measurements on the low temperature phase\cite{McDonald 2005} points
to the validity of electronic structure calculations for AuZn. From
the measured and calculated orbit of the high-temperature cubic
phase, Goddard \emph{et al}. concluded\cite{Goddard 2005} that the
Fermi surface is nested with a wave vector oriented along the (111)
direction of the cubic Brillouin zone that is nearly commensurate
($\sim {2 \over 3}$) with Brillouin zone and that the system lowers
its energy by undergoing a periodic modulation of the lattice which,
via electron-phonon coupling, opens a gap at the Fermi surface. The
acoustic de Haas-van Alphen effect supports this conclusion since it
not only reproduces some of the measured de Haas-van Alphen
frequencies but it is also direct experimental evidence of strong
electron-phonon interactions.

\section{Conclusions}

To summarize, we have shown that several materials show evidence
that the Fermi-surface electronic structure plays a significant role
in MTs. The MT in the bulk thermodynamical properties in nonmagnetic
alloys is significantly altered by the application of a magnetic
field. In the absence of ordered magnetism these results provide
strong evidence that the conduction electrons play an important role
in the MT mechanism. This conclusion is further confirmed by the
presence of quantum oscillations which represent a direct probe of
the Fermi surface. The cross-sectional frequencies observed here are
in favorable agreement with previously reported
observations~\cite{McDonald 2005,Goddard 2005}.

\begin{acknowledgments}
The authors would like to acknowledge Yongbin Lee, Bruce Harmon, and
Dan Thoma for useful discussions. The work at Temple was supported
by the United States Department of Energy, Office of Basic Energy
Sciences, through award DEFG02-01ER45872. The high magnetic field
measurements were carried out in Florida at the NHMFL under the
auspices of the United States Department of Energy, NSF, and the
State of Florida. Ultrasonic research at the NHMFL is supported by
the in-house research program. Work at Los Alamos was performed
under the auspices of the United States Department of Energy.
\end{acknowledgments}

\end{document}